\documentclass[sigconf, natbib=false]{acmart}

\usepackage[backend=biber, style=trad-plain]{biblatex}
\usepackage{comment, graphicx, hyperref, color, booktabs, xspace, algorithm, algpseudocode, amssymb, amsthm, subcaption}
\usepackage{mathtools}
\usepackage{balance} 

\newcommand{\name}{MagNet\xspace}

\DeclarePairedDelimiterX{\infdivx}[2]{(}{)}{%
  #1\;\delimsize\|\;#2%
}

\graphicspath{{fig/}}

\begin{document}

\title{\name: a Two-Pronged Defense against Adversarial Examples}

\author{Dongyu Meng}
\affiliation{%
  \institution{ShanghaiTech University}
}
\email{mengdy@shanghaitech.edu.cn}

\author{Hao Chen}
\affiliation{%
  \institution{University of California, Davis}
}
\email{chen@ucdavis.edu}

\begin{abstract}
Deep learning has shown impressive performance on hard perceptual problems. However, researchers found deep learning systems to be vulnerable to small, specially crafted perturbations that are imperceptible to humans. Such perturbations cause deep learning systems to mis-classify \emph{adversarial examples}, with potentially disastrous consequences where safety or security is crucial. Prior defenses against adversarial examples either targeted specific attacks or were shown to be ineffective.

We propose \name, a framework for defending neural network classifiers against adversarial examples. \name neither modifies the protected classifier nor requires knowledge of the process for generating adversarial examples. \name includes one or more separate detector networks and a reformer network. The detector networks learn to differentiate between normal and adversarial examples by approximating the manifold of normal examples. Since they assume no specific process for generating adversarial examples, they generalize well. The reformer network moves adversarial examples towards the manifold of normal examples, which is effective for correctly classifying adversarial examples with small perturbation. We discuss the intrinsic difficulties in defending against whitebox attack and propose a mechanism to defend against graybox attack. Inspired by the use of randomness in cryptography, we use diversity to strengthen \name. We show empirically that \name is effective against the most advanced state-of-the-art attacks in blackbox and graybox scenarios without sacrificing false positive rate on normal examples.

\end{abstract}


\copyrightyear{2017}
\acmYear{2017}
\setcopyright{acmlicensed}
\acmConference{CCS '17}{October 30-November 3, 2017}{Dallas, TX,
USA}\acmPrice{15.00}\acmDOI{10.1145/3133956.3134057}
\acmISBN{978-1-4503-4946-8/17/10}


\maketitle

\section{Introduction}
\label{sec:introduction}

In recent years, deep learning demonstrated impressive performance on many tasks, such as image classification~\cite{He:2016} and natural language processing~\cite{Kumar:2016}. However, recent research showed that an attacker could generate adversarial examples to fool classifiers~\cite{Szegedy:2014, Goodfellow:2015, Papernot:2016:Limitations, Liu:2017}. Their algorithms perturbed benign examples, which were correctly classified, by a small amount that did not affect human recognition but that caused neural networks to mis-classify. We call theses neural networks \emph{target classifiers}.

Current defenses against adversarial examples follow three approaches: (1) Training the target classifier with adversarial examples, called \emph{adversarial training}~\cite{Szegedy:2014, Goodfellow:2015}; (2) Training a classifier to distinguish between normal and adversarial examples~\cite{Metzen:2017}; and (3) 
Making target classifiers hard to attack by blocking gradient pathway, e.g., defensive distillation~\cite{Papernot:2016:Distillation}.

However, all these approaches have limitations. Both (1) and (2) require adversarial examples to train the defense, so the defense is specific to the process for generating those adversarial examples. For (3), Carlini et al.\ showed that defensive distillation did not significantly increase the robustness of neural networks~\cite{Carlini:2017}. Moreover, this approach requires changing and retraining the target classifier, which adds engineering complexities.

\begin{figure}[t]
\centering
\includegraphics[width=1.02\linewidth]{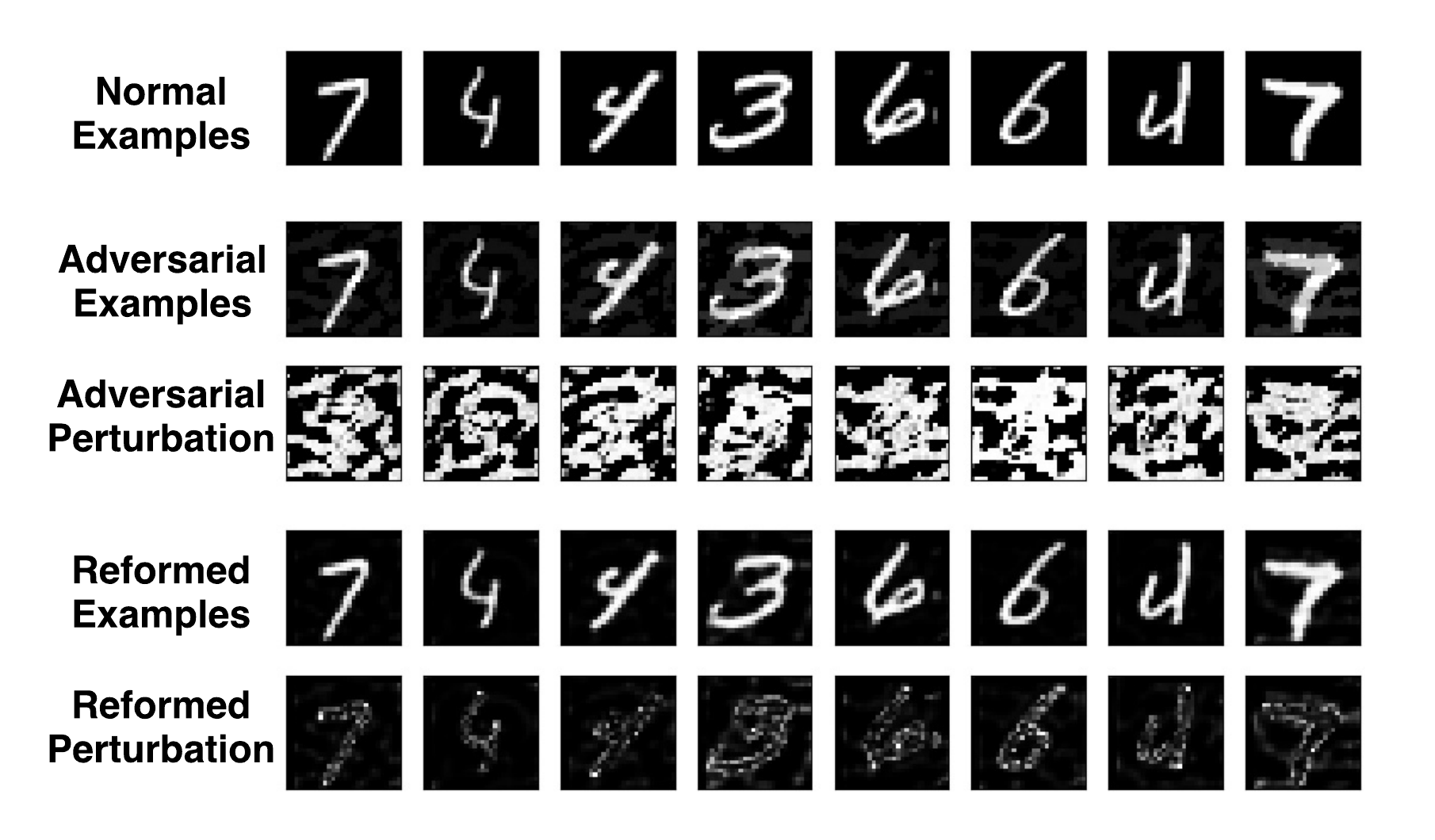}
\caption{An illustration of the reformer's effect on adversarial perturbations. The second row displays adversarial examples generated from the original normal examples in the first row by Carlini's $L^\infty$ attack. The third row shows their perturbations against the original examples, and these perturbations lack prominent patterns. The fourth row displays the adversarial examples after being reformed by \name. The fifth row displays the remaining perturbations in the reformed examples against their original examples in the first row, and these perturbations have the shapes of their original examples.}
\label{fig:diff}
\end{figure}

We propose \name\footnote{Imagine the manifold of normal examples as a magnet and test examples as iron particles in a high-dimensional space. The magnet is able to attract and move nearby particles (illustrating the effect of the \emph{reformer}) but is unable to move distant particles (illustrating the effect of the \emph{detectors}).}, a defense against adversarial examples with two novel properties. First, it neither modifies the target classifier nor relies on specific properties of the classifier, so it can be used to protect a wide range of neural networks. \name uses the target classifier as a blackbox: \name reads the output of the classifier's last layer, but neither reads data on any internal layer nor modifies the classifier. Second, \name is independent of the process for generating adversarial examples, as it requires only normal examples for training.

\subsection{Adversarial examples}

A \emph{normal example} $x$ for a classification task is an example that occurs naturally. In other words, the physical process for this classification task generates $x$ with non-negligible probability. For example, if the task is classifying handwritten digits, then the data generation process rarely generates an image of a tiger. An \emph{adversarial example} $y$ for a classifier is not a normal example and the classifier's decision on $y$ disagrees with human's prevailing judgment. See \autoref{sec:adversarialexamples} for a more detailed discussion. 

Researchers speculate that for many AI tasks, their relevant data lie on a manifold that is of much lower dimension than the full sample space~\cite{Narayanan:2010}. This suggests that the normal examples for a classification task are on a manifold, and adversarial examples are off the manifold with high probability.

\subsection{Causes of mis-classification and solutions}
\label{sec:misclassification}

A classifier mis-classifies an adversarial example for two reasons.

\begin{enumerate}
\item The adversarial example is far from the boundary of the manifold of the task. For example, the task is handwritten digit classification, and the adversarial example is an image containing no digit, but the classifier has no option to reject this example and is forced to output a class label.

\item The adversarial example is close to the boundary of the manifold. If the classifier generalizes poorly off the manifold in the vicinity of the adversarial example, then mis-classification occurs.
\end{enumerate}

We propose \name to mitigate these problems. To deal with the first problem, \name uses \emph{detectors} to detect how different a test example is from normal examples. A detector learns a function $f: \mathbb{X} \rightarrow \{0, 1\}$, where $\mathbb{X}$ is the set of all examples. $f(x)$ tries to measure the distance between the example $x$ and the manifold. If this distance is greater than a threshold, then the detector rejects $x$.

To deal with the second problem, \name uses a \emph{reformer} to reform adversarial examples. For this we use \emph{autoencoders}, which are neural networks trained to attempt to copy its input to its output. Autoencoders leverage simpler hidden representation to introduce regularization to uncover useful properties of the data~\cite{Goodfellow:2016, Vincent:2008, Vincent:2010}. We train an autoencoder with adequate normal examples for it to learn an approximate manifold of the data. Given an adversarial example $x$ close to the boundary of the manifold, we expect the autoencoder to output an example $y$ on the manifold where $y$ is close to $x$. This way, the autoencoder \emph{reforms} the adversarial example $x$ to a similar normal example $y$. \autoref{fig:diff} shows the effect of the reformer.

Since \name is independent of the target classifier, we assume that the attacker always knows the target classifier and its parameters. In the case of blackbox attack on \name, the attacker does not know the defense parameters. In this setting, we evaluated \name on popular attacks~\cite{Cleverhans:2016, Moosavi:2015, Carlini:2017}. On the MNIST dataset, \name achieved more than 99\% classification accuracy on adversarial examples generated by nine out of ten attacks considered. On the CIFAR-10 dataset, the classification accuracy improvement was also significant. Particularly, \name achieved high accuracy on adversarial examples generated by Carlini's attack, the most powerful attack known to us, across a wide range of confidence levels of the attack on both datasets. Note that we trained our defense without using any adversarial examples generated by the attack. In the case of whitebox attack, the attacker knows the parameters of \name. In this case, the attacker could view \name and the target classifier as a new composite classifier, and then generate adversarial examples against this composite classifier. Not surprisingly, we found that the performance of \name on whitebox attack degraded sharply. When we trained Carlini's attack on our reformer, the attack was able to generate adversarial examples that all fooled our reformer. In fact, we can view any defense against adversarial examples as enhancing the target classifier. As long as the enhanced classifier is imperfect (i.e., unable to match human decisions), adversarial examples are guaranteed to exist. One could make it difficult to find these examples, e.g., by hiding the defense mechanism or its parameters, but these are precluded in whitebox attack.

We advocate defense via diversity and draw inspiration from cryptography. The security of a good cipher relies on the diversity of its keys, as long as there is no better attack than searching the key space by brute force and this search is computationally infeasible. Adopting a similar approach, we create a number of different defenses and randomly pick one at run time. This way, we defend against \emph{graybox attack} (\autoref{sec:threatmodel}). In our implementation, we trained a number of different autoencoders as described above. If the attacker cannot predict which of these autoencoders is used at run time, then he has to generate adversarial examples that can fool all of them. As the diversity of these autoencoders grows, it becomes more difficult for the attacker to find adversarial examples. \autoref{sec:grayboxeval} will show that this technique raises the classification accuracy on Carlini's adversarial examples from 0 (whitebox attack) to 80\% (graybox attack).

We may also take advantage of these diverse autoencoders to build another detector, which distinguishes between normal and adversarial examples. The insight is that since normal examples are on the manifold, their classification decisions change little after being transformed by an autoencoder. By contrast, since adversarial examples are not on the manifold, their classification results change more significantly after being transformed by the autoencoder. We use the similarity between an example and its output from an autoencoder as a metric. But in contrast to the previous detector, which computes the distance between a test example and the manifold without consulting the target classifer, here we enlist the help from the target classifier. We assume that the classifier outputs the probability distribution of the test example on each label. Let this distribution be $p(y; x)$ for the original test example $x$, and $q(y; ae(x))$ for the output of the autoencoder $ae$ on $x$, where $y$ is the random variable for class labels. We use the Jensen-Shannon divergence between $p$ and $q$ as the similarity measure. Note that although this approach uses the target classifier, during training it does not depend on any specific classifier. It uses the classifier to compute the similarity measure only during testing. We found this detector more sensitive than the previous detector on powerful attacks (\autoref{sec:confidence}).

\subsection{Contributions}

We make the following contributions.

\begin{itemize}
\item We formally define adversarial example and metrics for evaluating defense against adversarial examples (\autoref{sec:adversarialexamples}).

\item We propose a defense against adversarial examples. The defense is independent of either the target classifier or the process for generating adversarial examples (\autoref{sec:detector}, \autoref{sec:reformer}).

\item We argue that it would be very difficult to defend against whitebox attacks. Therefore, we propose the graybox threat model and advocate defending against such attacks using diversity. We demonstrate our approach using diversity (\autoref{sec:grayboxdefense}).
\end{itemize}

\section{Background and related work}

\subsection{Deep learning systems in adversarial environments}

Deep learning systems play an increasingly important role in modern world. They are used in autonomous control for robots and vehicles~\cite{Bojarski:2016, Daftry:2016, Finn:2016}, financial systems~\cite{Sirignano:2016}, medical treatments~\cite{Shen:2017}, information security~\cite{Jung:2015, Pascanu:2015}, and human-computer interaction~\cite{Hinton:2012, Kahn:2017}. These security-critical domains require better understanding of neural networks from the security perspective.

Recent work has demonstrated the feasibility of attacking such systems with carefully crafted input for real-world systems~\cite{Carlini:2017, Papernot:2017, Grosse:2016}. More specifically, researchers showed that it was possible to generate adversarial examples to fool classifiers~\cite{Szegedy:2014, Goodfellow:2015, Papernot:2016:Limitations, Liu:2017}. Their algorithms perturbed normal examples by a small volume that did not affect human recognition but that caused mis-classification by the learning system. Therefore, how to protect such classifiers from adversarial examples is a real concern.

\subsection{Distance metrics}

By definition, adversarial examples and their normal counterparts should be visually indistinguishable by humans. Since it is hard to model human perception, researchers proposed three popular metrics to approximate human's perception of visual difference, namely $L^0$, $L^2$, and $L^\infty$~\cite{Carlini:2017}. These metrics are special cases of the $L^p$ norm:
\[
  \|x\|_p = \left(\sum_{i=1}^n|x_i|^p\right)^\frac{1}{p}
\]

These three metrics focus on different aspects of visual significance. $L^0$ counts the number of pixels with different values at corresponding positions in the two images. It answers the question of how many pixels are changed. $L^2$ measures the Euclidean distance between the two images. $L^\infty$ measures the maximum difference for all pixels at corresponding positions in the two images. 

Since there is no consensus on which metric is the best, we evaluated our defense on all these three metrics.

\subsection{Existing attacks}

Since the discovery of adversarial examples for neural networks in \cite{Szegedy:2014}, researchers have found adversarial examples on various network architectures. For example, feedforward convolutional classification networks~\cite{Carlini:2017}, generative networks~\cite{Kos:2017}, and recurrent networks~\cite{Papernot:2016:RNN}. These adversarial examples threaten a wide range of applications, e.g., classification\cite{Moosavi:2015} and semantic segmentation~\cite{Xie:2017}. Researchers developed several methods for generating adversarial examples, most of which leveraged gradient based optimization from normal examples~\cite{Carlini:2017, Szegedy:2014, Goodfellow:2015}. Moosavi et al.\ showed that it was even possible to find one effective universal adversarial perturbation that, when applied, turned many images adversarial~\cite{Moosavi:2016}.

To simplify the discussion, we only focus on attacks targeting neural network classifiers. We evaluated our defense against four popular, and arguably most advanced, attacks. We now explain these attacks.

\subsubsection{Fast gradient sign method(FGSM)}

Given a normal image $x$, fast gradient sign method~\cite{Goodfellow:2015} looks for a similar image $x'$ in the $L^\infty$ neighborhood of $x$ that fools the classifier. It defines a loss function $Loss(x, l)$ that describes the cost of classifying $x$ as label $l$. Then, it transforms the problem to maximizing $Loss(x', l_x)$ which is the cost of classifying image $x'$ as its ground truth label $l_x$ while keeping the perturbation small. Fast gradient sign method solves this optimization problem by performing one step gradient update from $x$ in the image space with volume $\epsilon$. The update step-width $\epsilon$ is identical for each pixel, and the update direction is determined by the sign of gradient at this pixel. Formally, the adversarial example $x'$ is calculated as:

\[
  x' = x + \epsilon \cdot sign(\nabla_xLoss(x, l_x))
\]

Although this attack is simple, it is fast and can be quite powerful. Normally, $\epsilon$ is set to be small. Increasing $\epsilon$ usually leads to higher attack success rate. For this paper, we use FGSM to refer to this attack.

\subsubsection{Iterative gradient sign Method}

\cite{Kurakin:2016} proposed to improve FGSM by using a finer iterative optimization strategy. For each iteration, the attack performs FGSM with a smaller step-width $\alpha$, and clips the updated result so that the updated image stays in the $\epsilon$ neighborhood of $x$. Such iteration is then repeated for several times. For the $ith$ iteration, the update process is:

\[
  x_{i+1}' = clip_{\epsilon, x}(x_i' + \alpha \cdot sign(\nabla_xLoss(x, l_x)))
\]

This update strategy can be used for both $L^\infty$ and $L^2$ metrics and greatly improves the success rate of FGSM attack. We refer to this attack as the iterative method for the rest of the paper.

\subsubsection{DeepFool}

DeepFool is also an iterative attack but formalizes the problem in a different way~\cite{Moosavi:2015}. The basic idea is to find the closest decision boundary from a normal image $x$ in the image space, and then to cross that boundary to fool the classifier. It is hard to solve this problem directly in the high-dimensional and highly non-linear space in neural networks. So instead, it iteratively solves this problem with a linearized approximation. More specifically, for each iteration, it linearizes the classifier around the intermediate $x'$ and derives an optimal update direction on this linearized model. It then updates $x'$ towards this direction by a small step $\alpha$. By repeating the linearize-update process until $x'$ crosses the decision boundary, the attack finds an adversarial example with small perturbation. We use the $L^\infty$ version of the DeepFool attack.

\subsubsection{Carlini attack}
\label{sec:bgcarlini}

Carlini recently introduced a powerful attack that generates adversarial examples with small perturbation~\cite{Carlini:2017}. The attack can be targeted or untargeted for all three metrics $L^0$, $L^2$, and $L^\infty$. We take the untargeted $L^2$ version as an example here to introduce its main idea. 

We may formalize the attack as the following optimization problem:

\[
\begin{aligned}
  & \underset{\delta}{\text{minimize}} & & \|\delta\|_2 + c \cdot f(x + \delta) \\
  & \text{such that} & & x+\delta \in [0, 1]^n
\end{aligned}
\]

For a fixed input image $x$, the attack looks for a perturbation $\delta$ that is small in length($\|\cdot\|$ term in objective) and fools the classifier(the $f(\cdot)$ term in objective) at the same time. $c$ is a hyper-parameter that balances the two. Also, the optimization has to satisfy the box constraints to be a valid image.

$f(\cdot)$ is designed in such a way that $f(x') \leqslant 0$ if and only if the classifier classifies $x'$ incorrectly, which indicates that the attack succeeds. $f(x')$ has hinge loss form and is defined as

\[
  f(x') = \text{max}(Z(x')_{l_x} - \text{max}\{Z(x')_i: i \neq l_x\}, -\kappa)
\]

where $Z(x')$ is the pre-softmax classification result vector (called logits) and $l_x$ is the ground truth label. $\kappa$ is a hyper-parameter called confidence. Higher confidence encourages the attack to search for adversarial examples that are stronger in classification confidence. High-confidence attacks often have larger perturbation and better transferability.

In this paper, we show that our defense is effective against Carlini's attack across a wide range of confidence levels (\autoref{sec:confidence}).

\subsection{Existing defense}

Defense on neural networks is much harder compared with attacks. We summarize some ideas of current approaches to defense and compare them to our work.

\subsubsection{Adversarial training}

One idea of defending against adversarial examples is to train a better classifier~\cite{Shaham:2015}. An intuitive way to build a robust classifier is to include adversarial information in the training process, which we refer to as adversarial training. For example, one may use a mixture of normal and adversarial examples in the training set for data augmentation~\cite{Szegedy:2014, Moosavi:2015}, or mix the adversarial objective with the classification objective as regularizer~\cite{Goodfellow:2015}. Though this idea is promising, it is hard to reason about what attacks to train on and how important the adversarial component should be. Currently, these questions are still unanswered.

Meanwhile, our approach is orthogonal to this branch of work. \name is an additional defense framework that does not require modification to the target classifier in any sense. The design and training of \name is independent from the target classifier, and is therefore faster and more flexible. \name may benefit from a robust target classifier (\autoref{sec:implementation}).

\subsubsection{Defensive distillation}

Defensive distillation \cite{Papernot:2016:Distillation} trains the classifier in a certain way such that it is nearly impossible for gradient based attacks to generate adversarial examples directly on the network. Defensive distillation leverages distillation training techniques~\cite{Hinton:2015} and hides the gradient between the pre-softmax layer (logits) and softmax outputs. However, \cite{Carlini:2017} showed that it is easy to bypass the defense by adopting one of the three following strategies: (1) choose a more proper loss function (2) calculate gradient directly from pre-softmax layer instead of from post-softmax layer (3) attack an easy-to-attack network first and then transfer to the distilled network.

We argue that in whitebox attack where the attacker knows the parameters of the defense network, it is very difficult to prevent adversaries from generating adversarial examples that defeat the defense. Instead, we propose to study defense in the graybox model (\autoref{sec:threatmodel}), where we introduce a randomization strategy to make it hard for the attacker to generate adversarial examples.

\subsubsection{Detecting adversarial examples}

Another idea of defense is to detect adversarial examples with hand-crafted statistical features~\cite{Grosse:2017} or separate classification networks~\cite{Metzen:2017}. An representative work of this idea is \cite{Metzen:2017}. For each attack generating method considered, it constructed a deep neural network classifier (detector) to tell whether an input is normal or adversarial. The detector was directly trained on both normal and adversarial examples. The detector showed good performance when the training and testing attack examples were generated from the same process and the perturbation was large enough, but it did not generalize well across different attack parameters and attack generation processes.

\name also employs one more more detectors. Contrary to previous work, however, we do not train our detectors on any adversarial examples. Instead, \name tries to learn the manifold of normal data and makes decision based on the relationship between a test example and the manifold. Further, \name includes a reformer that pushes hard-to-detect adversarial examples (with small perturbation) towards the manifold. Since \name is independent of any process for generating adversarial examples, it generalizes well.

\section{Problem definition}

\subsection{Adversarial examples}
\label{sec:adversarialexamples}
We define the following sets:

\begin{itemize}
\item $\mathbb{S}$: the set of all examples in the sample space (e.g., all images).

\item $\mathbb{C}_t$: the set of mutually exclusive classes for the classification task $t$. E.g., if $t$ is handwritten digit classification, then $\mathbb{C}=\{0, 1, \ldots, 9\}$. 

\item $\mathbb{N}_t = \{x | x \in \mathbb{S}$  and $x$ occurs naturally with regard to the classification task $t$ \}. Each classification task $t$ assumes a data generation process that generates each example $x \in \mathbb{S}$ with probability $p(x)$. $x$ occurs naturally if $p(x)$ is non-negligible. Researchers believe that $\mathbb{N}_t$ constitute a manifold that is of much lower dimension than $\mathbb{S}$~\cite{Narayanan:2010}. Since we do not know the data generation process, we approximate $\mathbb{N}_t$ by the union of natural datasets for $t$, such as CIFAR and MNIST for image recognition.
\end{itemize}

\begin{definition}
A \emph{classifier} for a task $t$ is a function $f_t: \mathbb{S} \rightarrow \mathbb{C}_t$
\end{definition}

\begin{definition}
  The \emph{ground-truth classifier} for a task $t$ represents human's prevailing judgment. We represent it by a function $g_t: \mathbb{S} \rightarrow \mathbb{C}_t \cup \{\bot\}$ where $\bot$ represents the judgment that the input $x$ is unlikely from $t$'s data generation process.
\end{definition}

\begin{definition}
An adversarial example $x$ for a task $t$ and a classifier $f_t$ is one where:
\begin{itemize}
\item $f_t(x) \not= g_t(x)$, and
\item $x \in \mathbb{S} \setminus \mathbb{N}_t$
\end{itemize}
\end{definition}

The first condition indicates that the classifier makes a mistake, but this in itself is not adequate for making the example adversarial. Since no classifier is perfect, there must exist natural examples that a classifier mis-classifies, so an attacker could try to find these examples. But these are not interesting adversarial examples for two reasons. First, traditionally they are considered as testing errors as they reflect poor generalization of the classifier. Second, finding these examples by brute force in large collections of natural examples is inefficient and laborious, because it would require humans to collect and label all the natural examples. Therefore, we add the second condition above to limit adversarial examples to only examples generated artificially by the attacker to fool the classifier.\footnote{Kurakin et al.\ showed that many adversarial images generated artificially remain adversarial after being printed and then captured by a camera~\cite{Kurakin:2016}. We still consider these as adversarial examples because although they occurred in physical forms, they were not generated by the natural process for generating normal examples.}

\subsection{Defense and evaluation}
\label{sec:defenseevaluation}

\begin{definition}
  A \emph{defense} against adversarial examples for a classifier $f_t$ is a function $d_{f_t}: \mathbb{S} \rightarrow \mathbb{C}_t \cup \{\bot\}$
\end{definition}

The defense $d_{f_t}$ extends the classifier $f_t$ to make it robust. The defense algorithm in $d_{f_t}$ may use $f_t$ in three different ways:

\begin{itemize}
\item The defense algorithm does not read data in $f_t$ or modify parameters in $f_t$.

\item The defense algorithm reads data in $f_t$ but does not modify parameters in $f_t$.

\item The defense algorithm modifies parameters in $f_t$.
\end{itemize}

When evaluating the effectiveness of a defense $d_{f_t}$, we cannot merely evaluate whether it classifies each example correctly, i.e., whether its decision agrees with that of the ground truth classifier $g_t$. After all, the goal of the defense is to improve the accuracy of the classifier on adversarial examples rather than on normal examples.

\begin{definition}
The defense $d_{f_t}$ makes a correct decision on an example $x$ if either of the following applies:

\begin{itemize}
\item $x$ is a normal example, and $d_{f_t}$ and the ground-truth classifier $g_t$ agree on $x$'s class, i.e., $x \in \mathbb{N}_t$ and $d_{f_t}(x)=g_t(x)$.

\item $x$ is an adversarial example, and either $d_{f_t}$ decides that $x$ is adversarial or that $d_{f_t}$ and the ground-truth classifier $g_t$ agree on $x$'s class, i.e., $x \in \mathbb{S} \setminus \mathbb{N}_t$ and ($d_{f_t}(x)=\bot$ or $d_{f_t}(x)=g_t(x))$.
\end{itemize}
\end{definition}

\subsection{Threat model}
\label{sec:threatmodel}

We assume that the attacker knows everything about the classifier $f_t$ that she wishes to attack, called \emph{target classifier}, such as its structure, parameters, and training procedure. Depending on whether the attacker knows the defense $d_{f_t}$, there are two scenarios:

\begin{itemize}
\item \emph{Blackbox attack}: the attacker does not know the parameters of $d_{f_t}$.

\item \emph{Whitebox attack}: the attacker knows the parameters of $d_{f_t}$.

\item \emph{Graybox attack}: except for the parameters, the attacker knows everything else about $d_{f_t}$, such as its structure, hyper-parameters, training set, training epochs. If we train a neural network multiple times while fixing these variables, we often get different model parameters each time because of random initialization. We can view that we get a different network each time. To push this one step further, we can train these different networks at the same time and force them to be sufficiently different by penalizing their resemblance. \autoref{sec:grayboxdefense} for an example. The defense can be trained with different structures and hyper-parameters for even greater diversity.

\end{itemize}

We assume that the defense knows nothing about how the attacker generates adversarial examples.

\section{Design}
\label{sec:design}

\begin{figure}[t]
\centering
\includegraphics[width=1\linewidth]{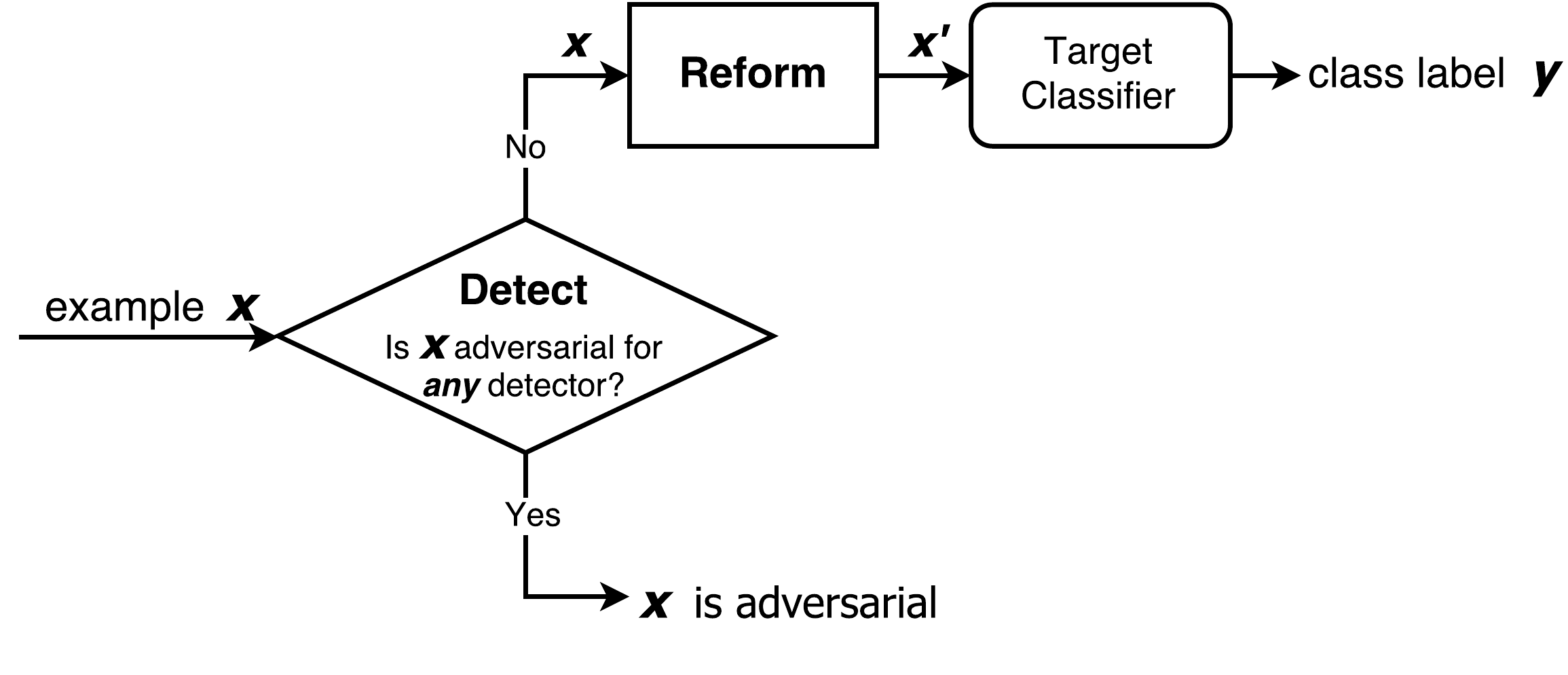}
  \caption{\name workflow in test phase. \name includes one or more detectors. It considers a test example $x$ adversarial if any detector considers $x$ adversarial. If $x$ is not considered adversarial, \name reforms it before feeding it to the target classifier.}
\label{fig:framework}
\end{figure}

\begin{figure}[t]
\centering
\includegraphics[width=1\linewidth]{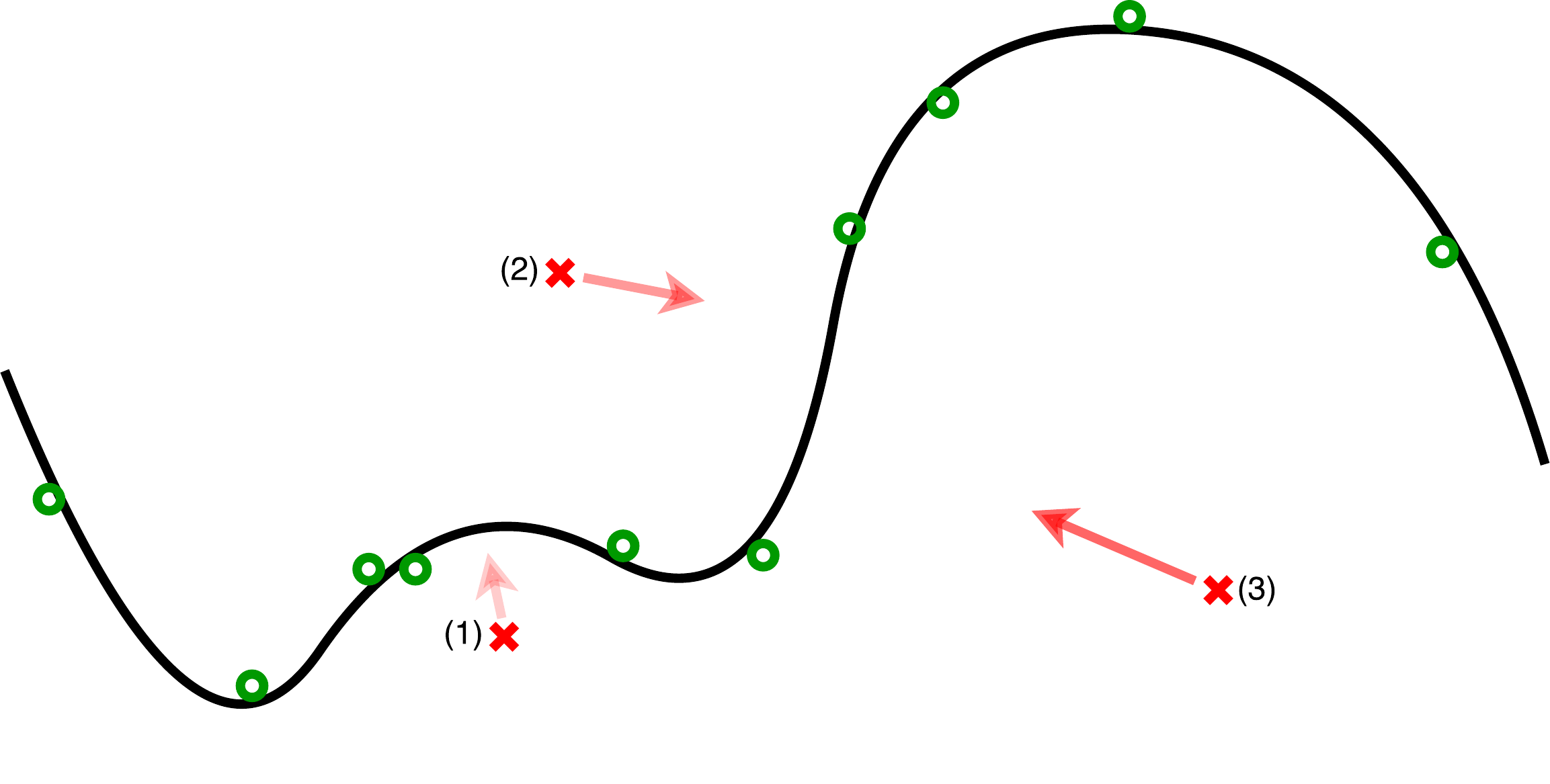}
  \caption{Illustration of how detector and reformer work in a 2-D sample space. We represent the manifold of normal examples by a curve, and depict normal and adversarial examples by green dots and red crosses, respectively. We depict the transformation by autoencoder using arrows. The detector measures reconstruction error and rejects examples with large reconstruction errors (e.g. cross (3) in the figure), and the reformer finds an example near the manifold that approximates the original example (e.g. cross (1) in the figure).}
\label{fig:manifold}
\end{figure}

\name is a framework for defending against adversarial examples (\autoref{fig:framework}). In \autoref{sec:misclassification} we provided two reasons why a classifier mis-classifies an adversarial example: (1) The example is far from the boundary of the manifold of normal examples, but the classifier has no option to reject it; (2) The example is close to the boundary of the manifold, but the classifier generalizes poorly off the manifold in the vicinity of the example. Motivated by these observations, \name consists of two components: (1) a \emph{detector} that rejects examples that are far from the manifold boundary, and (2) a \emph{reformer} that, given an example $x$, strives to find an example $x'$ on or close to the manifold where $x'$ is a close approximation to $x$, and then gives $x'$ to the target classifier. \autoref{fig:manifold} illustrates the effect of the detector and reformer in a 2-D sample space.

\subsection{Detector}
\label{sec:detector}

The detector is a function $d: \mathbb{S} \rightarrow \{0, 1\}$ that decides whether the input is adversarial. As an example of this approach, a recent work trained a classifier to distinguish between normal and adversarial examples~\cite{Metzen:2017}. However, it has the fundamental limitation that it requires the defender to model the attacker, by either acquiring adversarial examples or knowing the process for generating adversarial examples. Therefore, it unlikely generalizes to other processes for generating adversarial examples. For example, \cite{Metzen:2017} used a basic iterative attack based on the $L^2$ norm. Its results showed that if its detector was trained with slightly perturbed adversarial samples, the detector had high false positive rates because it decided many normal examples as adversarial. On the other hand, if the detector was trained with significantly perturbed examples, it would not be able to detect slightly perturbed adversarial examples.

\subsubsection{Detector based on reconstruction error}
\label{sec:detector-reconstruction}

To avoid the problem of requiring adversarial examples, \name's detector models only normal examples, and estimates the distance between the test example and boundary of the manifold of normal examples. Our implementation uses an autoencoder as the detector and uses the reconstruction error to approximate the distance between the input and the manifold of normal examples. An autoencoder $ae=d \circ e$ contains two components: an encoder $e: \mathbb{S} \rightarrow \mathbb{H}$ and a decoder $d: \mathbb{H} \rightarrow \mathbb{S}$, where $\mathbb{S}$ is the input space and $\mathbb{H}$ is the space of hidden representation. We train the autoencoder to minimize a loss function over the training set, where the loss function commonly is mean squared error: 
\[
L(\mathbb{X}_{\mathrm{train}})=\frac{1}{|\mathbb{X}_{\mathrm{train}}|} \sum_{x \in \mathbb{X}_{\mathrm{train}}}\|x - ae(x))\|_2
\]
The reconstruction error on a test example $x$ is 
\[
E(x)=\|x - ae(x))\|_p
\]

An autoencoder learns the features of the training set so that the encoder can encode the input with hidden representation of certain properties, and the decoder tries to reconstruct the input from the hidden representation. If an input is drawn from the same data generation process as the training set, then we expect a small reconstruction error. Otherwise, we expect a larger reconstruction error. Hence, we use reconstruction error to estimate how far a test example is from the manifold of normal examples. Since reconstruction error is a continuous value, we must set a threshold $t_{\mathrm{re}}$ for deciding whether the input is normal. This threshold is a hyper-parameter of an instance of detector. It should be as low as possible to detect slightly perturbed adversarial examples, but not too low to falsely flag normal examples. We decide $t_{\mathrm{re}}$ by a validation set containing normal examples, where we select the highest $t_{\mathrm{re}}$ such that the detector's false positive rate on the validation set is below a threshold $t_{\mathrm{fp}}$. This threshold $t_{\mathrm{fp}}$ should be decided catering for the requirement of the system.

When calculating reconstruction errors, it is important to choose suitable norms. Though reconstruction error based detectors are attack-independent, the norm choosen for detection do influence the sharpness of detection results. Intuitively, $p$-norm with larger $p$ is more sensitive to the maximum difference among all pixels, while smaller $p$ averages its concentration to each pixel. Empirically, we found it sufficient to use two reconstruction error based detectors with $L^1$ and $L^2$ norms respectively to cover both ends.

\subsubsection{Detector based on probability divergence}
\label{sec:detector-divergence}

The detector described in \autoref{sec:detector-reconstruction} is effective in detecting adversarial examples whose reconstruction errors are large. However, it becomes less effective on adversarial examples whose reconstruction errors are small. To overcome this problem, we take advantage of the target classifier.

Most neural network classifiers implement the softmax function at the last layer

\[
\mathrm{softmax}(\boldsymbol{l})_i=\frac{\mathrm{exp}(l_i)}{\sum_{j=1}^n\mathrm{exp}(l_j)}
\]

The output of softmax is a probability mass function over the classes. The input to softmax is a vector $\boldsymbol{l}$ called \emph{logit}. Let $\mathrm{rank}(\boldsymbol{l}, i)$ be the index of the element that is ranked the $i$th largest among all the elements in $\boldsymbol{l}$. Given a normal example whose logit is $\boldsymbol{l}$, the goal of the attacker is to perturb the example to get a new logit $\boldsymbol{l'}$ such that $\mathrm{rank}(\boldsymbol{l}, 1) \not= \mathrm{rank}(\boldsymbol{l'}, 1)$.

Let $f(x)$ be the output of the last layer (softmax) of the neural network $f$ on the input $x$. Let $ae(x)$ be the output of the autoencoder $ae$ that was trained on normal examples. If $x$ is a normal example, since $ae(x)$ is very close to $x$, the probability mass functions $f(x)$ and $f(ae(x))$ are similar. By contrast, if $x'$ is an adversarial example,  $ae(x')$ is significantly different from $x'$. We observed that even when the reconstruction error on $x'$ is small, $f(x')$ and $f(ae(x'))$ can be significantly different. This indicates that the divergence between $f(x)$ and $f(ae(x))$ reflects how likely $x$ is from the same data generation process as normal examples. We use Jensen-Shannon divergence:
\[
\mathrm{JSD}(P\parallel Q) = \frac{1}{2}D_{\mathrm{KL}}(P\parallel M) + \frac{1}{2}D_{\mathrm{KL}}(Q\parallel M)
\]
where
\[
D_{\mathrm{KL}}(P\parallel Q)=\sum_iP(i)\log \frac{P(i)}{Q(i)}
\]
and
\[
M=\frac{1}{2}(P+Q)
\]

When we implemented this, we encountered a numerical problem. Let $\boldsymbol{l}(x)$ be the logit of the input $x$. When the largest element in $\boldsymbol{l}(x)$ is much larger than its second largest element, softmax($\boldsymbol{l}(x)$) saturates, i.e., the largest element in softmax($\boldsymbol{l}(x)$) is very close to 1. When this happens, we observed that softmax($\boldsymbol{l}(ae(x))$) also saturates on the same element. This will make the Jensen-Shannon divergence between $\mathrm{softmax}(\boldsymbol{l}(x))$ and $\mathrm{softmax}(\boldsymbol{l}(ae(x)))$ very small. To overcome this numerical problem, we add a temperature $T>1$ when calculating softmax:

\[
\mathrm{softmax}(\boldsymbol{l})_i=\frac{\mathrm{exp}(l_i/T)}{\sum_{j=1}^n\mathrm{exp}(l_j/T)}
\]

\subsection{Reformer}
\label{sec:reformer}

The reformer is a function $r: \mathbb{S} \rightarrow \mathbb{N}_t$ that tries to reconstruct the test input. The output of the reformer is then fed to the target classifier. Note that we do not use the reformer when training the target classifier, but use the reformer only when deploying the target classifier. An ideal reformer:

\begin{enumerate}
\item should not change the classification results of normal examples.
\item should change adversarial examples adequately so that the reconstructed examples are close to normal examples. In other words, it should \emph{reform} adversarial examples.
\end{enumerate}

\subsubsection{Noise-based reformer}
A naive reformer is a function that adds random noise to the input. If we use Gaussian noise, we get the following reformer
\[
r(\boldsymbol{x}) = clip(\boldsymbol{x} + \epsilon \cdot \mathbf{y})
\]
where $\mathbf{y}$\textasciitilde$\mathcal{N}(\boldsymbol{y}; \boldsymbol{0}, \mathbf{I})$ is the normal distribution with zero mean and identity covariance matrix, $\epsilon$ scales the noise, and $clip$ is a function that clips each element of its input vector to be in the valid range.


A shortcoming of this noise-based reformer is that it fails to take advantage of the distribution of normal examples. Therefore, it changes both normal and adversarial examples randomly and blindly, but our ideal reformer should barely change normal examples but should move adversarial examples towards normal examples.

\subsubsection{Autoencoder-based reformer}
\label{sec:aereformer}

We propose to use autoencoders as the reformer. We train the autoencoder to minimize the reconstruction error on the training set and ensures that it generalizes well on the validation set. Afterwards, when given a normal example, which is from the same data generating process as the training examples, the autoencoder is expected to output a very similar example. But when given an adversarial example, the autoencoder is expected to output an example that approximates the adversarial example and that is closer to the manifold of the normal examples. In this way, \name improves the classification accuracy of adversarial examples while keeping the classification accuracy of normal examples unchanged.

\subsection{Use diversity to mitigate graybox attacks}
\label{sec:grayboxdefense}

In blackbox attacks, the attacker knows the parameters of the target classifier but not those of the detector or reformer. Our evaluation showed that \name was highly effective in defending against blackbox attacks (\autoref{sec:general}).

However, in whitebox attacks, where the attacker also knows the parameters of the detector and reformer, our evaluation showed that \name became less accurate. This is not surprising because we can view that \name transforms the target classifier $f_t$into a new classifier $f'_t$. In whitebox attacks, the attacker knows all the parameters of $f'_t$, so he can use the same method that he used on $f_t$ to find adversarial examples for $f'_t$. If such adversarial examples did not exist or were negligible, then it would mean that $f'_t$ agrees with the ground-truth classifier on almost all the examples off the manifold of normal example. Since there is no evidence that we could find this perfect classifier anytime soon, non-negligibly number of adversarial examples exist for any classifier, including $f'_t$. 

Although we cannot eliminate adversarial examples, we could make it difficult for attackers to find them. One approach would be to create a robust classifier such that even if the attacker knows all the parameters of the classifier, it would be difficult for her to find adversarial example~\cite{Papernot:2016:Distillation}. However, \cite{Carlini:2017} showed that it was actually easy to find adversarial examples for the classifier hardened in~\cite{Papernot:2016:Distillation}. We do not know how to find such robust classifiers, or even if they exist.

We take a different approach. We draw inspirations from cryptography, which uses randomness to make it computationally difficult for the attacker to find secrets, such as secret keys. We use the same idea to diversify our defense. In our implementation, we create a large number of autoencoders as candidate detectors and reformers. \name randomly picks one of these autoencoders for each defensive device for every session, every test set, or even every test example. Assume that the attacker cannot predict which autoencoder we pick for her adversarial example and that successful adversarial examples trained on one autoencoder succeed on another autoencoders with low probability, then the attacker would have to train her adversarial examples to work on all the autoencoders in our collection. We can increase the size and diversity of this collection to make the attack harder to perform. This way, we defend against graybox attack as defined in \autoref{sec:threatmodel}.

A key question is how to find large number of diverse autoencoders such that transfer attacks on target classifiers succeed with low probability. Rigorous theoretical analysis of the question is beyond the scope of this paper. Instead, we show a method for constructing these autoencoders and empirical evidence of its effectiveness.

We train $n$ autoencoders of the same or different architectures at the same time with random initialization. During training, in the cost function we add a regularization term to penalize the resemblance of these autoencoders
\begin{equation}
L(x) = \sum_{i=1}^n \mathrm{MSE}(x, ae_i(x)) - \alpha \sum_{i=1}^n \mathrm{MSE}(ae_i(x), \frac{1}{n}\sum_{j=1}^n ae_j(x))
\label{eqn:diversity}
\end{equation}
where $ae_i$ is the $i$th autoencoder, MSE is the mean squared error function, and $\alpha>0$ is a hyper-parameter that reflects the tradeoff between reconstruction error and autoencoder diversity. When $\alpha$ becomes larger, it encourages autoencoder diversity but also increases reconstruction error. We will evaluate this approach in \autoref{sec:grayboxeval}.

\section{Implementation and Evaluation}
\label{sec:implementation}

We evaluated the accuracy and properties of our defense described in \autoref{sec:design} on two standard dataset: MNIST~\cite{LeCun:1998} and CIFAR-10~\cite{Krizhevsky:2009}.

\subsection{Setup}
\label{sec:setup}

On MNIST, we selected 55\,000 examples for the training set, 5\,000 for the validation set, and 1\,000 for the test set. We trained a classifier using the setting in \cite{Carlini:2017} and got an accuracy of 99.4\%. On CIFAR-10, we selected 45\,000 examples for training set, 5\,000 for the validation set, and 10\,000 for the test set. We used the architecture in \cite{Springenberg:2014} and got an accuracy of 90.6\%. The accuracy of both these classifiers is near the state of the art on these datasets. \autoref{tbl:archs} and \autoref{tbl:hyper}
show the architecture and training parameters of these classifiers. We used a scaled range of [0, 1] instead of [0, 255] for simplicity.

\begin{table}[t] 
\caption{Architecture of the classifiers to be protected}
\begin{center}
\begin{tabular}{lllll}
\toprule
MNIST & & & CIFAR & \\
\midrule
Conv.ReLU & 3 $\times$ 3 $\times$ 32 & &  Conv.ReLU & 3 $\times$ 3 $\times$ 96\\
Conv.ReLU & 3 $\times$ 3 $\times$ 32 & &  Conv.ReLU & 3 $\times$ 3 $\times$ 96\\
Max Pooling & 2 $\times$ 2           & &  Conv.ReLU & 3 $\times$ 3 $\times$ 96\\
Conv.ReLU & 3 $\times$ 3 $\times$ 64 & &  Max Pooling & 2 $\times$ 2 \\
Conv.ReLU & 3 $\times$ 3 $\times$ 64 & &  Conv.ReLU & 3 $\times$ 3 $\times$ 192\\
Max Pooling & 2 $\times$ 2           & &  Conv.ReLU & 3 $\times$ 3 $\times$ 192\\
Dense.ReLU & 200                     & &  Conv.ReLU & 3 $\times$ 3 $\times$ 192\\
Dense.ReLU & 200                     & &  Max Pooling & 2 $\times$ 2 \\
Softmax & 10                         & &  Conv.ReLU & 3 $\times$ 3 $\times$ 192\\
        &                            & &  Conv.ReLU & 1 $\times$ 1 $\times$ 192\\
        &                            & &  Conv.ReLU & 1 $\times$ 1 $\times$ 10\\
        &                            & &  \multicolumn{2}{l}{Global Average Pooling}  \\
        &                            & &  Softmax & 10\\
\bottomrule
\end{tabular}
\end{center}
\label{tbl:archs}
\end{table}

\begin{table}[t] 
\caption{Training parameters of classifiers to be protected}
\begin{center}
\begin{tabular}{lll}
\toprule
Parameters & MNIST & CIFAR \\
\midrule
Optimization Method & SGD       & SGD   \\
Learning Rate       & 0.01      & 0.01  \\
Batch Size          & 128       & 32    \\
Epochs              & 50        & 350   \\
Data Augmentation   & -       & Shifting + Horizontal Flip \\
\bottomrule
\end{tabular}
\end{center}
\label{tbl:hyper}
\end{table}

In the rest of this section, first we evaluate the robustness of \name in blackbox attack, where the attacker does not know the parameters used in \name. To understand why \name works and when it works well, we analyze the impact of the detector and the reformer, respectively, on the accuracy of \name against Carlini's attack. Finally, we evaluate the use of diversity to mitigate graybox attack, where we use the same classifier architecture but train it to get many classifiers of different parameters.

We may divide attacks using adversarial examples into two types.  In \emph{targeted attack}, the attacker chooses a particular class and then creates adversarial examples that the victim classifier mis-classifies into that class. In \emph{untargeted attack}, the attacker does not care which class the victim classifier outputs as long as it is different from the ground truth. Previous work showed that untargeted attack is easier to succeed, results in smaller perturbations, and transfers better to different models~\cite{Liu:2017, Carlini:2017}. Since untargeted attack is more difficult to defend against, we evaluate \name on untargeted attack to show its worst case performance.

\subsection{Overall performance against blackbox attacks}
\label{sec:general}

We tested \name against attacks using fast gradient sign method, iterative gradient sign method, DeepFool, and Carlini's method. For fast gradient sign method and iterative gradient sign method, we used the implementation of Cleverhans~\cite{Cleverhans:2016}. For DeepFool and Carlini's attack, we used their authors' open source implementations~\cite{Moosavi:2015, Carlini:2017}. 

In principle, \name works better when we deploy several instances of both reconstruction error based detectors and probability divergence based detectors. Diversified autoencoder architecture also boosts defense performance. In our implementation, we try to simplify the setup by limiting our detector usage and sharing architectures among autoencoders. This is for convenience rather than mandatory. More specifically, for MNIST dataset, we only use two reconstruction error based detectors of two unique architectures. For CIFAR-10 dataset, we share the same structure among all autoencoders. \autoref{tbl:defarchs_mnist}, \autoref{tbl:defarchs_cifar}, and \autoref{tbl:defhyper} show the architectures and training hyper-parameters of the autoencoder for MNIST and CIFAR-10. We tune the network to make sure it works, but make no further effort to optimize these settings.

\begin{table}[t] 
\caption{Defensive devices architectures used for MNIST, including both encoders and decoders.}
\begin{center}
\begin{tabular}{lllll}
\toprule
  \multicolumn{2}{l}{Detector I \& Reformer} & & Detector II & \\
\midrule
  Conv.Sigmoid   & 3 $\times$ 3 $\times$ 3   & & Conv.Sigmoid & 3 $\times$ 3 $\times$ 3 \\
  AveragePooling & 2 $\times$ 2              & & Conv.Sigmoid & 3 $\times$ 3 $\times$ 3 \\
  Conv.Sigmoid   & 3 $\times$ 3 $\times$ 3   & & Conv.Sigmoid & 3 $\times$ 3 $\times$ 1 \\
  Conv.Sigmoid   & 3 $\times$ 3 $\times$ 3   & & & \\
  Upsampling     & 2 $\times$ 2              & & & \\
  Conv.Sigmoid   & 3 $\times$ 3 $\times$ 3   & & & \\
  Conv.Sigmoid   & 3 $\times$ 3 $\times$ 1   & & & \\
\bottomrule
\end{tabular}
\end{center}
\label{tbl:defarchs_mnist}
\end{table}

\begin{table}[t] 
\caption{Defensive devices architecture used for CIFAR-10, including both encoders and decoders.}
\begin{center}
\begin{tabular}{lllll}
\toprule
  \multicolumn{2}{l}{Detectors \& Reformer} \\
\midrule
  Conv.Sigmoid & 3 $\times$ 3 $\times$ 3 \\
  Conv.Sigmoid & 3 $\times$ 3 $\times$ 3 \\
  Conv.Sigmoid & 3 $\times$ 3 $\times$ 1 \\
\bottomrule
\end{tabular}
\end{center}
\label{tbl:defarchs_cifar}
\end{table}

\begin{table}[t] 
\caption{Training parameters for defensive devices.}
\begin{center}
\begin{tabular}{lll}
\toprule
Parameters & MNIST & CIFAR \\
\midrule
Optimization Method & Adam            & Adam                 \\
Learning Rate       & 0.001           & 0.001                \\
Batch Size          & 256             & 256                  \\
Epochs              & 100             & 400                  \\
Regularization      & $L^2(10^{-9})$  & Noise                \\ 
\bottomrule
\end{tabular}
\end{center}
\label{tbl:defhyper}
\end{table}

Below we use the criteria described and justified in \autoref{sec:defenseevaluation} to evaluate the accuracy of \name on normal and adversarial examples.

\subsubsection{MNIST}

Compared to CIFAR-10, MNIST is an easier dataset for classifiers. We trained a classifier to achieve an accuracy of 99.4\%, which is close to the state of the art. We found that we needed only the reconstruction error-based detector and reformer to become highly accurate against adversarial examples generated from MNIST. Therefore, we did not include the probability divergence-based detector in \name in this evaluation. Detector II and detector I(see \autoref{tbl:defarchs_mnist}) used the $L^2$ and $L^1$ norm to measure reconstruction error, respectively. We selected the threshold of reconstruction error such that the false positive rate of the detector on the validation set is at most 0.001, i.e., each detector mistakenly rejects no more than 0.1\% examples in the validation set.

\paragraph{Effect on normal examples} On the test set, without \name, the accuracy of the classifier is 99.4\%; with \name, the accuracy is reduced to 99.1\%. This small reduction is negligible.

\paragraph{Effect on adversarial examples} \autoref{tbl:mnistoverall} shows that the accuracy of \name is above 99\% on all the attacks considered except Carlini attack with $L^0$ norm(92.0\%). Note that we achieved such high accuracy without training \name on any of these attacks.

\begin{table}
\caption{Classification accuracy of \name on adversarial examples generated by different attack methods. Some of these attacks have different parameters on MNIST and CIFAR-10 because they need to adjust their parameters according to datasets.}
\begin{subtable}{\linewidth} 
\caption{MNIST}
\begin{center}
\begin{tabular}{lllll}
\toprule
  Attack      & Norm                 & Parameter                  & No Defense   & With Defense \\
\midrule                                                                                
  FGSM        & $L^\infty $          & $\epsilon=0.005$           & 96.8\%       & 100.0\%      \\       
  FGSM        & $L^\infty $          & $\epsilon=0.010$           & 91.1\%       & 100.0\%      \\       
  Iterative   & $L^\infty $          & $\epsilon=0.005$           & 95.2\%       & 100.0\%      \\       
  Iterative   & $L^\infty $          & $\epsilon=0.010$           & 72.0\%       & 100.0\%      \\       
  Iterative   & $L^2      $          & $\epsilon=0.5$             & 86.7\%       & 99.2\%       \\       
  Iterative   & $L^2      $          & $\epsilon=1.0$             & 76.6\%       & 100.0\%      \\        
  Deepfool    & $L^\infty $          &                            & 19.1\%       & 99.4\%       \\       
  Carlini     & $L^2      $          &                            & 0.0\%        & 99.5\%       \\        
  Carlini     & $L^\infty $          &                            & 0.0\%        & 99.8\%       \\        
  Carlini     & $L^0      $          &                            & 0.0\%        & 92.0\%       \\        
\bottomrule 
\end{tabular}
\end{center}
\label{tbl:mnistoverall}
\end{subtable}

\vspace{1ex}

\begin{subtable}{\linewidth}
\caption{CIFAR}
\begin{center}
\begin{tabular}{lllll}
\toprule
  Attack      & Norm         & Parameter                   & No Defense   & With Defense \\
\midrule 
  FGSM        & $L^\infty$   & $ \epsilon=0.025$           & 46.0\%       & 99.9\%       \\
  FGSM        & $L^\infty$   & $ \epsilon=0.050$           & 40.5\%       & 100.0\%      \\
  Iterative   & $L^\infty$   & $ \epsilon=0.010$           & 28.6\%       & 96.0\%       \\
  Iterative   & $L^\infty$   & $ \epsilon=0.025$           & 11.1\%       & 99.9\%       \\
  Iterative   & $L^2     $   & $ \epsilon=0.25$            & 18.4\%       & 76.3\%       \\
  Iterative   & $L^2     $   & $ \epsilon=0.50$            & 6.6\%        & 83.3\%       \\
  Deepfool    & $L^\infty$   &                             & 4.5\%        & 93.4\%       \\
  Carlini     & $L^2     $   &                             & 0.0\%        & 93.7\%       \\ 
  Carlini     & $L^\infty$   &                             & 0.0\%        & 83.0\%       \\ 
  Carlini     & $L^0     $   &                             & 0.0\%        & 77.5\%       \\ 
\bottomrule 
\end{tabular}
\end{center}
\label{tbl:cifaroverall}
\end{subtable}
\label{tbl:overall}
\end{table}

\subsubsection{CIFAR-10}

Compared to MNIST, CIFAR-10 is more challenging for classifiers. We trained a classifier to achieve an accuracy of 90.6\%, which is close to the state of the art. For CIFAR-10, \name consists of a reformer, a reconstruction error-based detector, and two probability divergence-based detectors with temperature $T$ of 10 and 40, respectively. We trained the autoencoder as a denoising autoencoder with unit Gaussian noise with volume 0.025. Error based detector uses the $L^1$ norm to measure reconstruction error. Again, we set a threshold of false positive rate $t_{\mathrm{fp}}$ on the validation set to find the threshold of reconstruction error. We set $t_{\mathrm{fp}}$ to 0.005 for the reconstruction error-based detector, and 0.01 for the probability divergence-based detector.

\paragraph{Effect on normal examples} 
On the test set, without \name, the accuracy of the classifier is 90.6\%; with \name, the accuracy is reduced to 86.8\%. The reduction in accuracy is small.

\paragraph{Effect on adversarial examples}
\autoref{tbl:cifaroverall} shows that the accuracy of \name on 10 different attacks. \name is not as accurate on CIFAR-10 as on MNIST, because the target classifier is not as strong on CIFAR-10 and leaves less space for \name to take effect. \name achieved an accuracy above 75\% on all the attacks, and above 90\% on more than half attacks. This provides empirical evidence that \name is effective and generalizes well to different attacks and different parameters of the same attack.

\subsection{Case study on Carlini attack, why does \name work?}
\label{sec:confidence}

Carlini showed that it was viable to mount transfer attack with higher confidence on MNIST~\cite{Carlini:2017}. Among the attacks that we evaluated, Carlini's attack is the most interesting because it is the most effective on the distillation defense~\cite{Papernot:2016:Distillation} and there is no known effective defense prior to our work. This attack is also interesting because the attacker can change the attack strength by adjusting the confidence level when generating adversarial examples. The higher confidence is, the stronger classification confidence is, and the larger distortion gets. At a confidence level of 40, the attack achieved a success rate of close to 100\% on classifier with distillation defense even by conducting \emph{transfer attack}.

We evaluated the impact of different confidence levels in Carlini's attack on \name. For MNIST, we used the same classifier as in Carlini's paper~\cite{Carlini:2017} for generating adversarial examples and as the target classifier in our evaluation. We generated adversarial examples with confidence levels in the range of [0, 40]. For CIFAR-10, \cite{Carlini:2017} did not evaluate the impact of confidence level, but we picked confidence levels in the range of [0, 100]. We use the classifier in \autoref{sec:general} for CIFAR-10 as target classifier. We keep the defense setting in \autoref{sec:general} unchanged for both datasets.

\begin{figure}
\centering
\includegraphics[width=1.1\linewidth]{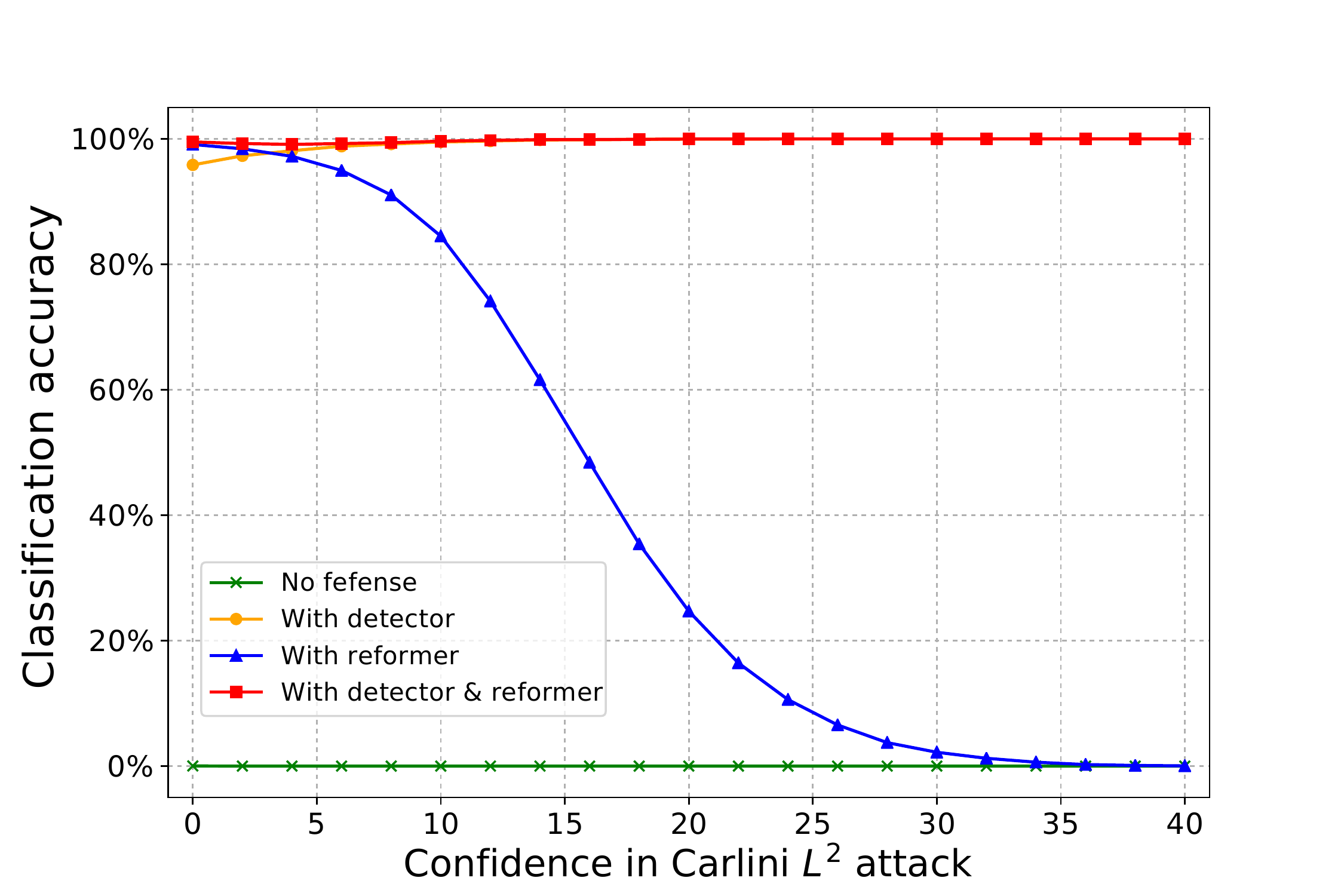}
\caption{Defense performance with different confidence of Carlini's $L^2$ attack on MNIST dataset. The performance is measured as the percentage of adversarial examples that are either detected by the detector, or classified correctly by the classifier.}
\label{fig:mnistconfidence}
\end{figure}

\autoref{fig:mnistconfidence} shows the performance of the detector and reformer on MNIST. Without \name, the attack succeeded almost 100\%, i.e., the classification accuracy rate is close to 0. With \name, the classification accuracy rate is above 99\% on adversarial examples generated at all confidence levels tested. This indicates that \name blocks Carlini attack completely in blackbox scenario.

\begin{figure}
\centering
\includegraphics[width=1.1\linewidth]{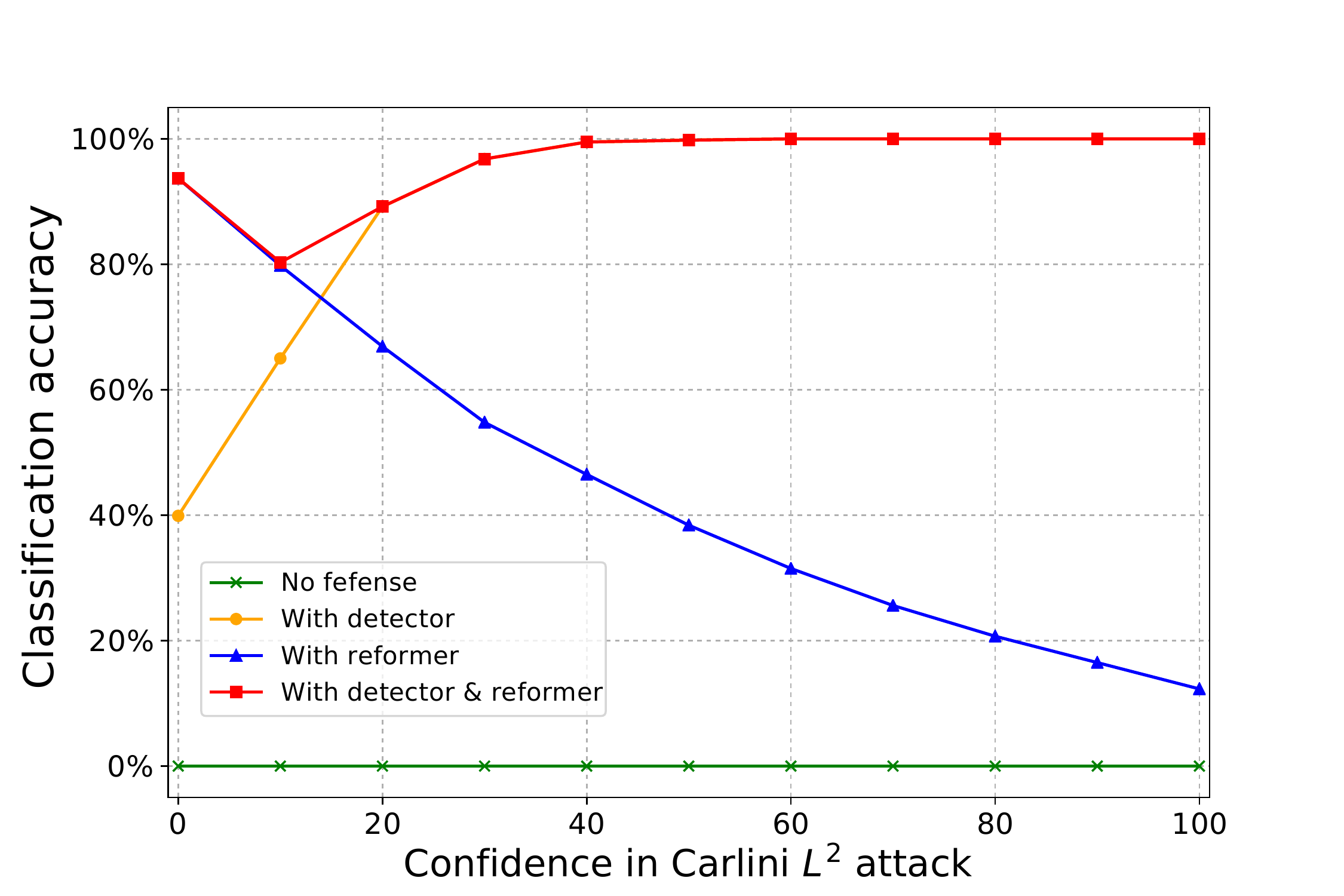}
\caption{Defense performance on different confidence of Carlini's $L^2$ attack on CIFAR-10 dataset. The performance is measured as the percentage of adversarial examples that are either detected by the detector, or classified correctly by the classifier.}
\label{fig:cifar_confidence_all}
\end{figure}

\autoref{fig:cifar_confidence_all} shows the classification accuracy of \name on CIFAR-10. The attack also gets near 100\% success rate for all confidences. A striking revelation in \autoref{fig:cifar_confidence_all} is that the detector and reformer compensate each other to achieve an overall high accuracy at all confidence levels. At high confidence level, the adversarial example is far from the manifold of normal examples, so it likely has a high reconstruction error, and therefore will be rejected by the detector. At low confidence level, the adversarial example is close to the manifold of normal examples, so the reconstructed example by the reformer is more likely to lie on the manifold and therefore to be classified correctly. In other words, as the confidence level of the adversarial example goes up, the reformer becomes less effective but the detector becomes more effective, so there is a dip in the mid range on the curve of the overall classification accuracy as shown in \autoref{fig:cifar_confidence_all}. This dip is an window of opportunity for the attacker, as it is where the effectiveness of the reformer begins to wane but the power of detectors have not started. In \autoref{fig:cifar_confidence_all}, even though this window of opportunity exists, \name still achieves classification accuracy above 80\% at all confidence levels.

Same dip should have appeared in \autoref{fig:mnistconfidence}, but the classifier and \name is strong enough to fill the dip.

\begin{figure}
\centering
\includegraphics[width=1.1\linewidth]{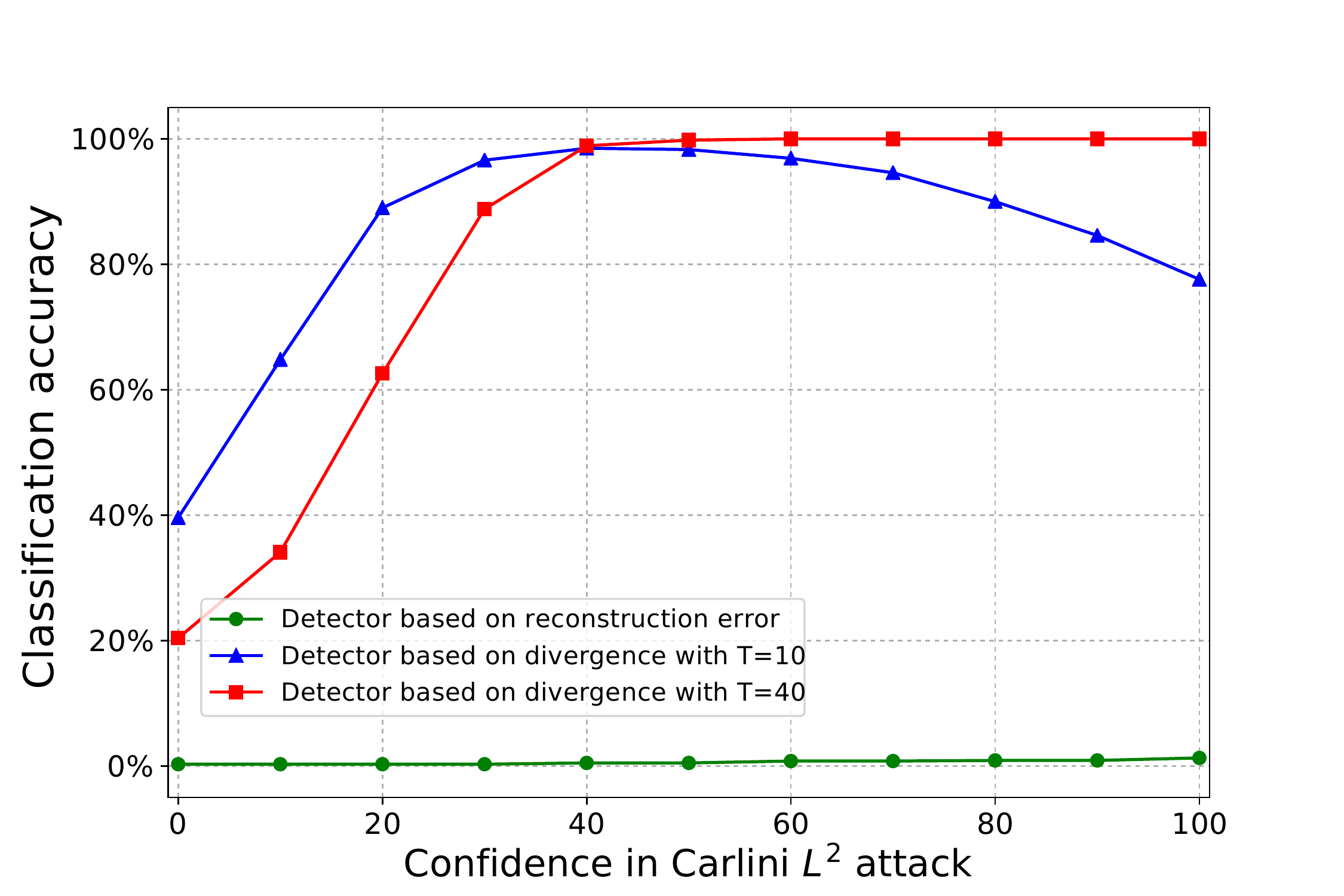}
\caption{Defense performance on different confidence of Carlini's $L^2$ attack on CIFAR-10 dataset. The performance is measured as the percentage of adversarial examples that are either detected by the detector, or classified correctly by the classifier.}
\label{fig:cifar_confidence_det}
\end{figure}

\autoref{fig:cifar_confidence_det} shows the effect of the temperature $T$ on the accuracy of the probability divergence-based detector. Low temperature makes the detector more accurate on adversarial examples at low confidence level, and high temperature makes the detector more accurate on adversarial examples at high confidence level.

Note again that we did not train \name with Carlini's attack or any other attacks, so we conjecture that the results likely generalize to other attacks.

\subsection{Defend against graybox attacks}
\label{sec:grayboxeval}

In graybox attack, except for the parameters, the attacker knows everything else about the defense, such as network structure, training set, and training procedure. If we assume that (1) the attacker cannot predict the parameters that the defender uses when classifying her adversarial examples; and (2) the attacker cannot feasibly mislead all possible defense when generating her adversarial examples, then we can defend against attackers by diversifying our defensive network.

We show an example defense against graybox attack. In this example, we provide diversity by training $n$ different autoencoders for the reformer in \name. In our proof-of-concept implementation, we used the same architecture, a convolutional autoencoder with $3 \times 3 \times 8$ hidden layers and ReLU activation, to obtain eight autoencoders of different parameters. During training, we used the same hyper-parameters as in \autoref{sec:general} except that we first trained the eight autoencoders independently for 3 epochs using the standard mean squared error loss. Then, we continued training these autoencoders using the loss in \autoref{eqn:diversity} for another 10 epochs, where we chose $\alpha=0.2$ empirically. At test time, we randomly picked one of the eight autoencoders as the reformer.

We chose Carlini's attack to evaluate this defense. However, Carlini's attack models only one network and uses the decision of the network to decide how to perturb the candidate adversarial example. But \name contains at least two networks, a reformer and one (or more) detector, that make independent decisions. Therefore, the attack as described in~\cite{Carlini:2017} cannot handle \name. To overcome this obstacle, we removed the detectors from \name and kept only the reformer to allow Carlini's attack to generate adversarial examples. But in this case, it would not fair to test \name with adversarial examples at high confidence level, because \name relies on the detector to reject adversarial examples at high confidence level (\autoref{fig:cifar_confidence_all}). Therefore, we ran Carlini attack to generate adversarial examples at confidence level 0. We chose only CIFAR-10 because Carlini's attack is more effective on it than on MNIST.

\autoref{tbl:diversityadversarial} shows the classification accuracy of \name on adversarial examples generated by Carlini's attack. We name each autoencoder A through H. Each column corresponds to an autoencoder that the attack is generated on, and each row corresponds to an autoencoder that is used during testing. The last row, \emph{random}, means that \name picks a random one from its eight autoencoders. The diagonal shows that \name's classification accuracy drops to mostly 0 when the autoencoder on which Carlini's attack was trained is also the one that \name used during testing. However, when these two autoencoders differ, the classification accuracy jumps to above 90\%. The last row shows a more realistic scenario when the attacker chooses a random autoencoder during training and \name also chooses a random autoencoder during testing from the eight candidate autoencoders. In this case, \name maintains classification accuracy above 80\%.

\begin{table}[t] 
\caption{Classification accuracy in percentage on adversarial examples generated by graybox attack on CIFAR-10. We name each autoencoder A through H. Each column corresponds to an autoencoder that the attack is trained on, and each row corresponds to an autoencoder that is used during testing. The last row, \emph{random}, means that \name picks a random one from its eight autoencoders.}
\begin{center}
\begin{tabular}{lllllllll}
\toprule
                  &  A       &  B     &  C     &  D     &  E      &  F     &  G      &  H       \\
\midrule 
    A             &    0.0   &  92.8  &  92.5  &  93.1  &  91.8   &  91.8  &   92.5  &   93.6 \\
    B             &    92.1  &  0.0   &  92.0  &  92.5  &  91.4   &  92.5  &   91.3  &   92.5 \\
    C             &    93.2  &  93.8  &  0.0   &  92.8  &  93.3   &  94.1  &   92.7  &   93.6 \\  
    D             &    92.8  &  92.2  &  91.3  &  0.0   &  91.7   &  92.8  &   91.2  &   93.9 \\ 
    E             &    93.3  &  94.0  &  93.4  &  93.2  &  0.0    &  93.4  &   91.0  &   92.8 \\
    F             &    92.8  &  93.1  &  93.2  &  93.6  &  92.2   &  0.0   &   92.8  &   93.8 \\
    G             &    92.5  &  93.1  &  92.0  &  92.2  &  90.5   &  93.5  &   0.1   &   93.4 \\
    H             &    92.3  &  92.0  &  91.8  &  92.6  &  91.4   &  92.3  &   92.4  &   0.0  \\
    Random   &    81.1  & 81.4   & 80.8   & 81.3   & 80.3    & 81.3   & 80.5    & 81.7 \\
\bottomrule 
\end{tabular}
\end{center}
\label{tbl:diversityadversarial}
\end{table}

\autoref{tbl:diversitynormal} shows the classifier accuracy of these autoencoders on the test set for CIFAR-10. Compared to the accuracy of the target classifier, 90.6\%, these autoencoders barely reduce the accuracy of the target classifier.

\begin{table}[t] 
\caption{Classification accuracy in percentage on the test set for CIFAR-10. Each column corresponds to a different autoencoder chosen during testing. ``Rand'' means that \name randomly chooses an autoencoder during testing.}
\begin{center}
\begin{tabular}{@{}llllllllll@{}}
\toprule
AE &A       &  B     &  C     &  D     &  E      &  F     &  G      &  H   & Rand    \\
\midrule 
Acc & 89.2 & 88.7 & 89.0 & 89.0 & 88.7 & 89.3 & 89.2 & 89.1 & 89.0\\
\bottomrule 
\end{tabular}
\end{center}
\label{tbl:diversitynormal}
\end{table}

There is much room for improvement on how to diversify \name. We could use autoencoders of different architectures, tune autoencoders with different training parameters, increase the amount of autoencoders, and encourage the difference between these autoencoders. We leave these for future work.

\section{Discussion}
\label{sec:discussion}

The effectiveness of \name against adversarial examples depends on the following assumptions:

\begin{itemize}
\item There exist detector functions that measure the distance between its input and the manifold of normal examples.

\item There exist reformer functions that output an example $x'$ that is perceptibly close to the input $x$, and $x'$ is closer to the manifold than $x$.
\end{itemize}

We chose autoencoder for both the reformer and the two types of detectors in \name. \name's high accuracy against the state-of-the-art attacks provides empirical evidence that our assumptions are likely correct. However, before we find stronger justification or proof, we cannot dismiss the possibility that our good results occurred because the state-of-the-art attacks are not powerful enough. We hope that our results would motivate further research on finding more powerful attacks or more powerful detectors and reformers.

\section{Conclusion}

We proposed \name, a framework for defending against adversarial perturbation of examples for neural networks. \name handles untrusted input using two methods. It detects adversarial examples with large perturbation using detector networks, and pushes examples with small perturbation towards the manifold of normal examples. These two methods work jointly to enhance the classification accuracy. Moreover, by using autoencoder as detector networks, \name learns to detect adversarial examples without requiring either adversarial examples or the knowledge of the process for generating them, which leads to better generalization. Experiments show that \name defended against the state-of-art attacks effectively. In case that the attacker knows the training examples of \name, we described a new graybox threat model and used diversity to defend against this attack effectively.

We advocate that defense against adversarial examples should be attack-independent. Instead of finding properties of adversarial examples from specific generation processes, a defense would be more transferable by finding intrinsic common properties among all adversarial generation processes. \name is a first step towards this end and demonstrated good performance empirically.

\begin{acks}
We thank Dr.\ Xuming He and anonymous reviewers for their valuable feedback.
\end{acks}

\printbibliography
\balance
\end{document}